\documentclass[10pt,twocolumn,showpacs,aps,prl]{revtex4}
\usepackage{amsmath}
\usepackage{amsfonts}
\usepackage{amssymb}
\usepackage{graphicx}
\usepackage{caption}
\usepackage{dcolumn}
\usepackage{latexsym}
\begin {document}


\title {Stable target opinion through power law bias in information exchange}
\author
{Amitava Datta$^1$}
\email{amitava.datta@uwa.edu.au}
\affiliation
{
\begin {tabular}{c}
$^1$Department of Computer Science and Software Engineering, University of Western Australia, Perth, WA 6009, Australia 
\end{tabular}
}
\begin{abstract}
We study a model of binary decision making when a certain population of agents 
is initially seeded with 
two different opinions, `$+$' and `$-$', with fractions $p_1$ and 
$p_2$ respectively, $p_1+p_2=1$. 
Individuals can reverse their initial opinion only once 
based on this information exchange. We study this model on a completely connected network, where 
any pair of agents can exchange information, and a two-dimensional square lattice with 
periodic boundary conditions, where 
information exchange is possible only between the nearest neighbors. We propose a model 
in which each agent maintains two counters of opposite opinions and accepts opinions 
of other agents with a power law bias until a threshold is reached, when they fix their final opinion. 
Our model is inspired by the study of negativity bias and positive-negative asymmetry known in the 
psychology literature for a long time. Our model can achieve stable intermediate mix of positive and 
negative opinions in a population. In particular,
we show that it is possible to achieve close to any fraction $p_3, 0\leq p_3\leq 1$,
of `$-$' opinion 
starting from an initial fraction $p_1$ of `$-$' opinion by applying a bias through adjusting the power law 
exponent of $p_3$.    
\end{abstract}

\maketitle

\section {I. Introduction}

The study of emergent behaviour in a population based on simple interaction 
rules among individuals has been an intense area of research in complex systems 
and sociophysics. Most of these models study opinion formation 
in a population in the context of two different opinions, majority and minority, 
or `$+$' and `$-$'. 
The opinions in these models evolve either according to some simple local rules, or according 
to some group dynamics. An extensive review of such models can be found in the paper by
Castellano {\em et al.}\cite{CFL}.

The aim of this paper is to investigate {\em negativity bias} in opinion formation. There is 
an extensive literature on negativity bias in psychology, as reviewed by Rozin and Royzman \cite{RR} and Vaish {\em et al.}
\cite{VGW}. Negativity bias 
is manifested in humans and animals in many different activities, including, attention and salience, 
sensation and perception, motivation, mood and decision making \cite{RR}. Some of these activities are closely related to 
opinion formation 
and hence, it is interesting to study the effect of negativity bias in opinion formation in a population. We present 
a model of opinion formation that uses negativity bias and has several interesting properties, including a 
similarity with the random-field Ising model and also the formation of predictable intermediate 
configurations of mixed opinions.  

One of the earliest among opinion formation models was the voter 
model (VM) \cite{L,LI} that can be simulated on any connected network.  
Each agent has a state $\pm 1$ and two neighboring agents interact at each simulation 
time step. Starting from a ($+-$) state, the probablity of assuming a ($++$) 
or a ($--$) state in an interaction is $\frac{1}{2}$ each. This simple update 
rule gives rise to rich dynamical behavior and the VM has been studied extensively. 
The VM always evolves to a homogeneous final state of one of the opinions, 
the rate of convergence depends on the initial populations of the two opinions, and has 
a stochastic nature. Hence it is hard to predict the mix of populations at intermediate 
stages of the evolution. 

Schweitzer and Behera \cite{SB} introduced the nonlinear VM where neighbors with different 
opinions are weighted with nonlinear weighting factors. The nonlinear VM has interesting 
configurations where both the opinions coexist equally when the starting initial fraction of population for 
each opinion is $0.5$ of the total. 
However it is not clear whether the nonlinear VM has any stable intermediate 
configurations where both opinions coexist if the initial populations are unbalanced. 

A {\em contrarian} has a different strategy from the other agents. 
Contrarians introduce interesting variations in the evolution of almost all the models we discuss. 
Masuda \cite{M} studied the linear VM by introducing three types of contrarians and concluded that 
contrarians prevent the evolution of the linear VM to a homogeneous final state of a single opinion 
and induce a mixed population of both the opinions. Masuda \cite{M} derived the equilibrium distributions of 
the two opinions under different assumptions on the contrarians. However it is not clear 
whether it is possible to get specific mix of populations in the model in \cite{M} and also 
whether the dynamics of the model is scale invariant. 

Group dynamics of binary opinion formation has been studied by Galam and his coworkers 
extensively as discussed in the review paper \cite{G}. 
Galam's \cite{G1} 2-state opinion dynamics model is of particular interest 
for our present work. This model is for a completely connected network where any agent is a 
neighbor of any other agent. Initially each agent has one of two opinions A and B, and the density of the 
two populations is denoted by $p_c$, expressed as a fraction, e.g., $p_c=\frac{1}{2}$ indicates a balanced initial population 
of agents with A and B opinions. 
Each step of the evolution of the model consists of picking a random group of agents of a predermined size. All agents in this 
group adopt the opinion of the local majority. When this update process is repeated, the resulting dynamics 
is dependent on $p_c$. The final population is a balance of A and B opinions when $p_c=\frac{1}{2}$ and the group size 
is odd. When the group size is even, the final population is balanced for a different value of $p_c$, however, 
any deviation from these $p_c$ values 
makes the final population to converge to one of the two opinions. 
The speed of convergence is faster for larger group sizes. 

Galam \cite{G1} also considered the introduction of contrarian agents in this model, the contrarians 
participate in the group opinion formation exactly in the same way as before, however, a contrarian
reverses its opinion once it has left the group. A mixed phase dynamics with a clear separation of 
majority and minority opinions  prevails when 
the density of the contrarians $a_c$ is low. These populations are stable for a fixed value of $a_c$.   
However, there are thresholds for $a_c$ for different group sizes when no opinion 
dominates and there is no symmetry breaking to separate the final population into majority and minority opinions. 
In other words the final population is balanced between the two opinions even though agents change their 
opinion dynamically. It is not clear whether Galam's model can achieve arbitrary and stable proportions 
of the two opinions by introducing contrarians. 

The Majority Rule (MR) model was introduced by Krapivsky and Redner \cite{KR} as a simple two-state opinion 
dynamics model. The MR model has similarities with Galam's model \cite{G1}. A group of agents is chosen at every 
step and the agents in the group all assume the opinion of the local majority. The aim in \cite{KR} was to study the
time to reach global consensus as a function of $N$, the total number of agents and also the probability of 
reaching a given final state as a function of the initial opinion densities. The MR model has many interesting 
properties and one of the characteristics features 
of the MR model that is of interest to us  is that even small islands of one opinion surrounded by the opposite
opinion can grow in size. The growth of a particular opinion varies from one initialization to another. 
There are also intermediate metastable states in the MR model that persist for long times, however again the 
concentration of opinions in these metastable states vary depending on the initialization.   

There are  some similarities between our proposed model and the model in \cite{G1}. First, 
the aim of our model is to arrive at a final population of a mixed majority and minortiy population. This is 
achieved in Galam's model when the density of the contrarians is low. Second, our model behaves similar to 
Galam's model when the fraction of initial population of agents is balanced, i.e., $\frac{1}{2}$ each. This is 
manifested in Galam's model both in the absence of contrarians and also when the initial population of 
contrarians is greater than a threshold for different group sizes. 

However there are distinct dissimilarities between our model and the model in \cite{G1}, apart 
from the fact that the update rules in Galam's model are based on groups. Our aim is to 
achieve a final population of agents separated into majority and minority opinions without the 
use of contrarians. In other words, all agents in our model have a common strategy. Galam's model without contrarians 
has been analysed for $d$-dimensional lattices by Lanchier and Taylor \cite{LT}. They have proven 
that Galam's model (or the {\em spatial public debate model} in the terminology of \cite{LT}) converges to a 
stationary distribution where both the opinions have positive densities. However it is not clear whether 
any specific mix of the two opinions can be achieved.  
Our model on the 
other hand behaves in a similar way both on a completely connected network and on a 2D lattice with 
nearest neighbor connections. Hence our model can be thought of as the formation of global opinion through simple local 
interactions. The differnces between our model and the MR model are also similar to the differences mentioned above.  

We frame our problem in this paper in a general way
as follows. Given an initial population of agents with `$+$'  and `$-$' opinions, with fractions 
$p_1$ and $p_2$ respectively, $p_1+p_2=1$, the goal is to achieve a 
fraction $p_3$ of final population 
of agents with the `$-$'  opinion, $0\leq p_3\leq 1$, and $p_3>p_2$. 
We show that it is possible to achieve close to a final fraction 
$p_3$ of agents with `$-$' opinion by introducing a bias in the exchange of opinion 
when two agents meet. It is interesting that this bias can be expressed as a power law 
exponent of $p_3$, and scale-invariant for both the completely 
connected network and the two dimensional lattice. Our model has interesting properties 
that are similar to other models studied in statistical mechanics. For example, a coexistence 
of opposite opinions has been observed in the nonlinear voter model \cite{SB}, even though 
this coexistence is not stable and predictable in terms of the exact mix of the two opinions. Also 
properties of our model related to the surface tension of the domain boundaries of opposite
opinions and also first-order phase transition and domain formation are similar to the random-field Ising model \cite{CN}. 

The rest of the paper is organized as follows. We discuss our model in section II. 
We discuss the results from an empirical study of the model without and with the 
power-law bias during information exchange in sections III and IV respectively. 
Finally we conclude in section V.

\section{II. The model}

A population initially has agents with two opinions in certain fractions 
$p_1$ and $p_2$, with $p_1+p_2=1$. Each agent has the option of choosing one of the 
opinions as their final opinion, however, reverting an initial opinion is allowed only 
once. Agents interact pairwise either on a completely connected network or on a square 
lattice with periodic boundary conditions. We have two free parameters in our model, 
$\beta$ and $\tau$. Each agent maintains two counters $\theta^+$
and $\theta^-$ of the positive and negative opinions encountered so far. Initially 
$\theta^+_i=1$ and $\theta^-_i=0$ if agent $i$ is a `$+$' agent, and 
$\theta^+_i=0$ and $\theta^-_i=1$ if agent $i$ is a `$-$' agent. When agents 
$i$ and $j$ interact, the rules for exchange of opinion from 
agent $i$ to agent $j$ are given in eqn.(\ref{eqn1}) and eqn.(\ref{eqn2}) 
(the subscripts
of the two counters indicate which agent the counter belongs to): 

\begin{equation}
\label{eqn1}
if\;\; (\theta^-_i> p_3^{\beta}\theta^+_i)\;\;then\;\; \theta^-_j=\theta^-_j+1
\end{equation}

\begin{equation}
\label{eqn2}
if\;\; (\theta^+_i>\theta^-_i)\;\;then\;\; \theta^+_j=\theta^+_j+1
\end{equation} 

The update of the state of agent $j$ occurs due to one of these two equations in a Monte Carlo step. 
First eqn.(\ref{eqn1}) is checked and if it is satisfied and an update occurs, eqn.(\ref{eqn2}) is 
skipped for that Monte Carlo step. Otherwise the condition in eqn.(\ref{eqn2}) is checked and 
an update occurs if the condition in eqn.(\ref{eqn2}) is satisfied.   
There is no exchange of information if $\theta^+_i=\theta^-_i$ and the 
encounter is considered a failure. In other words, we introduce an asymmetry in the 
updating of the counters of agent $j$ by introducing the bias factor 
$p_3^\beta$ in eqn.(\ref{eqn1}). The other free parameter $\tau$ is 
used as a threshold of opinion until which an agent participates in information exchange. 
Once either of the counters $\theta^+_i$ or $\theta^-_i$ reaches $\tau$, agent $i$ freezes
its state either to `$+$', or to `$-$', depending on whether $\theta^+_i>\theta^-_i$ or 
$\theta^+_i<\theta^-_i$. Hence this freezing of state may require a flipping of the initial 
state of agent $i$, this is allowed only once. Once frozen, the state of agent $i$ remains the 
same until all agents have reached the threshold which is same for all agents. However 
agent $i$ still maintains its two counters $\theta^+_i$ and $\theta^-_i$ for 
use in interactions with agents who have not yet reached the threshold $\tau$. 
Though there are no contrarian agents in our model, the $\beta$ parameter can be viewed as similar to a 
contrarian strategy as it introduces a bias in the interactions among the agents. 

The hallmark of negativity bias is to give greater emphasis to negative perceptions and entities. 
This emphasis is manifested in four different ways \cite{RR}, negative potency, steeper negative gradients, negativity dominance 
and negative differentiation. We aim to capture negative potency and steeper negative gradient in our 
model. Negative potency gives a stronger impact for negative entities, compared to positive entities. This is 
captured in eqn ({\ref{eqn1}). Negative gradient emphasises the steeper growth of negative events compared to 
positive events. This is an outcome of our model as we will explain in later sections. It becomes harder 
for positive opinion to prevail as negative opinion accummulates more and more in the 
counters of the agents in our model. The use of a single exponent in the power law bias for the 
whole population tries to capture the inherent negativity bias quantitatively through a single 
exponent. Though this is a simplified assumption in eqn.(\ref{eqn1}), we show later that the behavior of 
the system is quite stable when this exponent is allowed to vary randomly within a certain range. 

The agents in our model maintain more information compared to Galam's model \cite{G1} and  the MR model \cite{KR}
in the sense that an agent has access to both accumulated 
positive and negative opinions of another agent during an interaction. It may seem that we 
are assuming a lot more information for decision making,  
However one surprising aspect of our model is that the convergence to the desired final 
state is  very fast. In other words, 
each agent needs to interact with a much smaller number of other agents in order to 
arrive at a `correct' decision, so that the overall fraction of desired opinion 
is achieved. Also our model can converge to a desired `$-$' opinion above the $0.5$ threshold
accurately even when the initial fraction of agents with `$-$' opinion is low. This is not the case 
for all the models that we have reviewed above. Galam's model \cite{G1} can achieve such a 
target population only by using contrarian agents. The MR model has intermediate states with 
mixed populations, however the mix is sensitive to instances of the two initial populations 
even when the fraction of the two initial populations are fixed. 

We will show in the following sections that our model has stable states of mix of opinions that 
can be tuned fairly accurately using the $\beta$ and $\tau$ parameters. These stable states are scale-invariant both 
for the completely connected network and the 2D lattice. However our model is unstable with respect to the $\tau$ parameter. 
The final configurations converge to a single opinion if the threshold $\tau$ is set relatively high. This 
convergence is faster  on the completely connected network and much slower on the 2D lattice. However, 
the parameter $\tau$ is a measure of the number of interactions among the agents and a lower value 
indicates that the convergence of our model to a balanced population mix is faster.  

Forming opinion based on accumulated history may be a realistic model  
in the sense that people in real life accept others' opinions for decision making. Also 
people have their own opinion and they usually take others' opinions with some negativity bias.
The contagious effects of negative opinions has been studied in the psychology literature \cite{RR}, 
and it has been noted that negative aspects of human thought process spread faster \cite{RN}.  
Moreover people discussing a binary 
opinion usually talk about the pros and cons of the two choices. Though it is hard to 
capture these processes through some numerical estimates, agents in our model abstract such real-life
interactions and discussions through the counters and also through the bias for accepting opinions.

\begin{figure}
\centering
\includegraphics[width=\columnwidth,height=6cm]{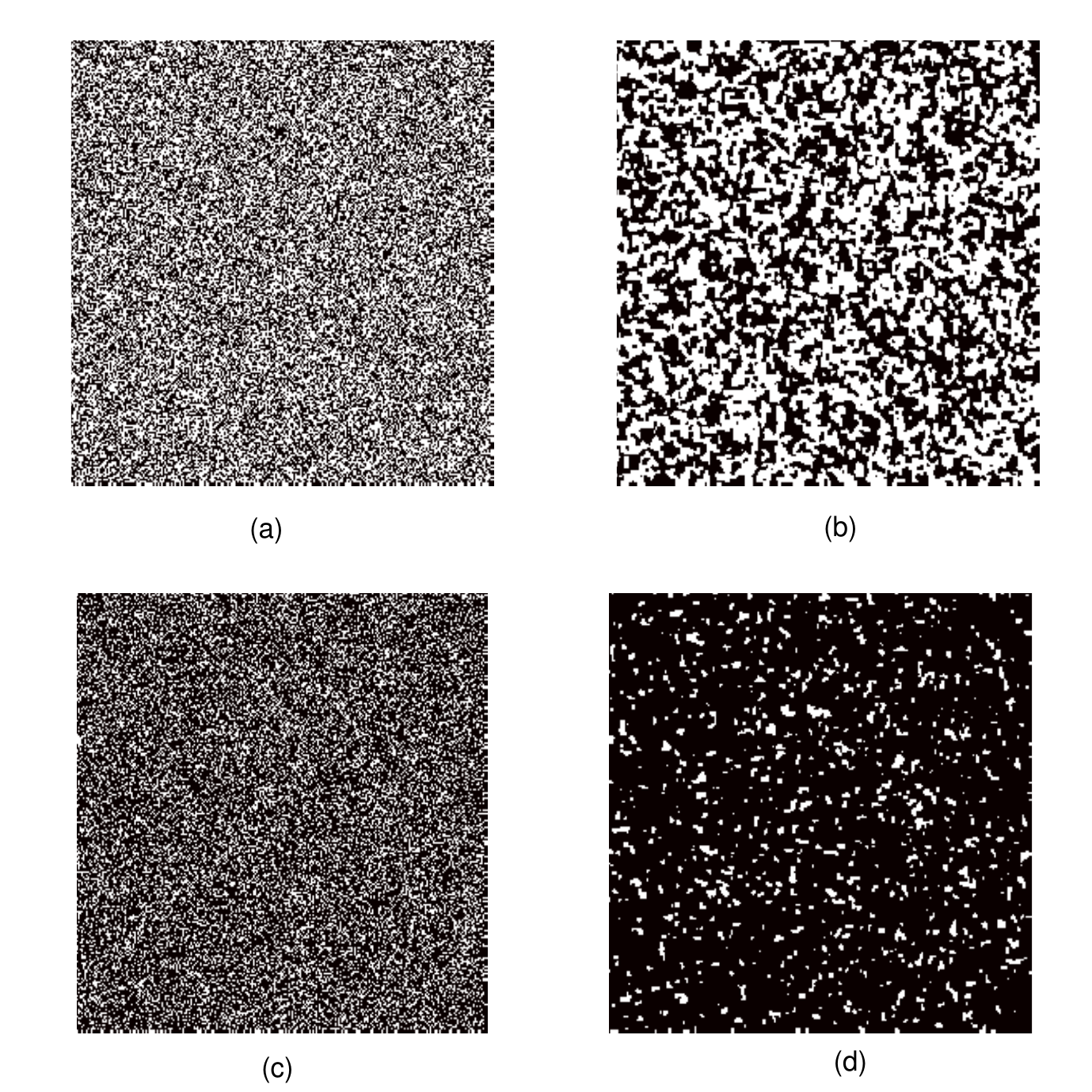}
\caption{\small{Two simulation results on a $256\times 256$ lattice, `$-$' agents are shown in black and 
`$+$' agents in white, simulation time ($t$) is measured in Monte Carlo (MC) steps.
{\bf (a)} Initial
fractions of populations: `$+$' and `$-$' both $0.5$ of the total; {\bf (b)} The lattice after all agents
reach the threshold $\tau=10$, at $t=2657346$. The final population of `$-$' agents
was a fraction of $0.51$ of the total; 
{\bf (c)} Initial fractions of populations: `$-$'opinion $0.7$ and
`$+$' opinion $0.3$; {\bf (d)} The lattice after all agents reach the threshold $\tau=10$, at $t=2500178$. The final population of `$-$' agents 
was a fraction of $0.9$ of the total. 
}}
\label{fig1}
\end{figure}

\begin{figure}
\centering
\includegraphics[width=\columnwidth,height=8.5cm]{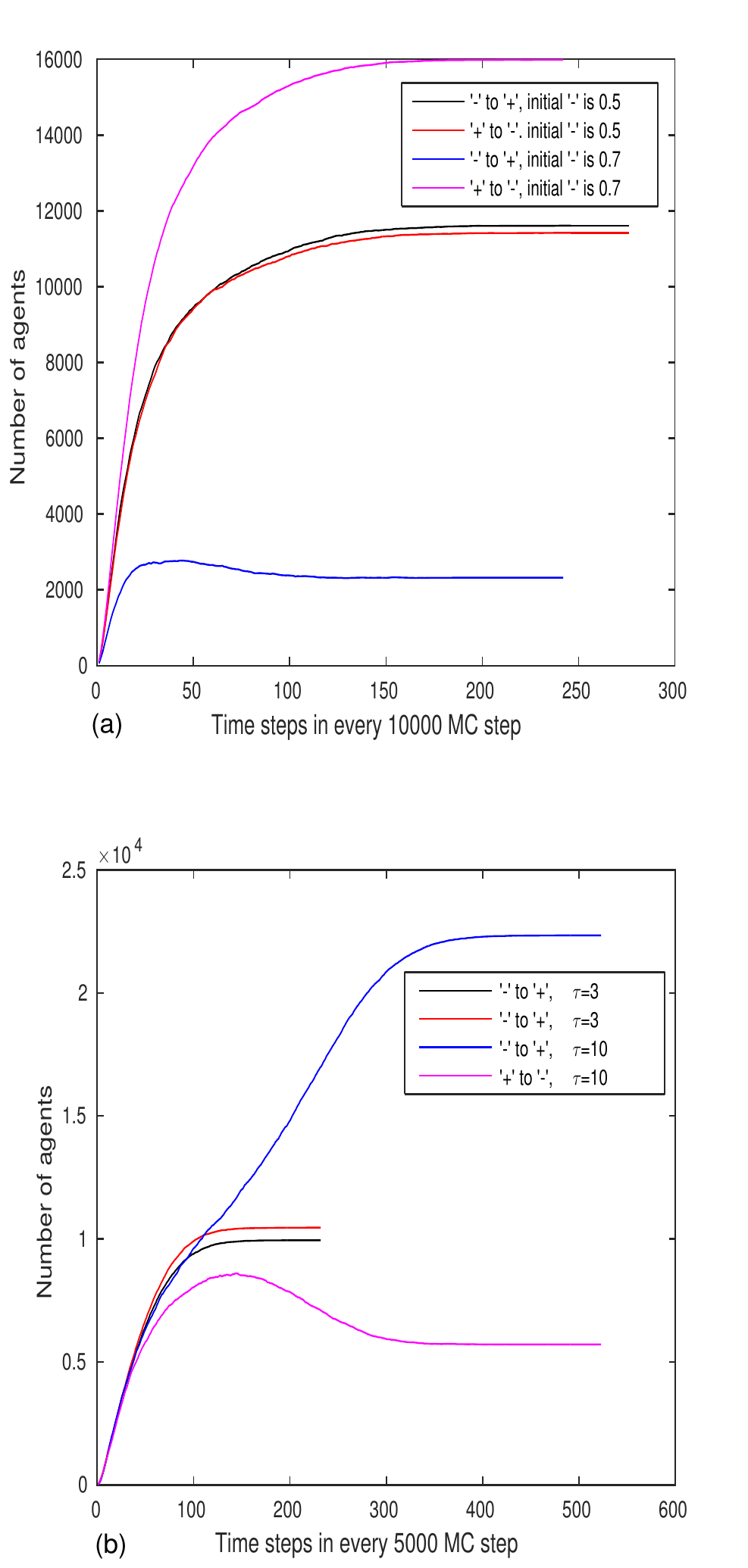}
\caption{\small{The conversion of agents from `$+$' to `$-$' and `$-$' to `$+$'. {\bf (a)} On a 
$256\times 256$ lattice, corresponding to the simulations in Fig. \ref{fig1}. The middle two plots are for the simulation 
when the initial population is balanced, $0.5$ each of `$+$' and `$-$' agents. The top and bottom plots are for 
an initial population of `$-$' agents $0.7$ of the total. The conversion of `$+$' to `$-$' is much higher (top plot), 
compared to the conversion of `$-$' to `$+$' (bottom plot). {\bf (b)} On a 
completely connected network (CCN). The two middle plots are for a small threshold $\tau=3$, 
and the conversions are similar. The conversions vary widely from run to run (bottom and top plots) when the threshold is 
higher ($\tau=10$).
}}
\label{figg1}
\end{figure}

\section{III. Dynamics without bias}

We first discuss the dynamics of our model without applying any bias, in other words when 
$\beta=0$ in eqn.(\ref{eqn1}). There is no bias for `$+$' or `$-$'  
opinion in this case, and agent $j$ accepts the higher of the two counters
$\theta^+_i$ and $\theta^-_i$ of agent $i$ for updating its own counter. We study 
some interesting dynamics of our model on a $256\times 256$ lattice in Fig. \ref{fig1}. 
We have verified that these results are scale-invariant by simulating 
them on lattices of size $512\times 512$ and $1024\times 1024$. 

Fig.\ref{fig1} shows results of two simulations of the unbiased system on a $256\times 256$ 
lattice with wrap-around connections. Panels (a) and (b) show the results when the initial populations of `$+$' and `$-$' agents 
are a fraction of $0.5$ each, and the threshold is $\tau=10$. We have simulated this configuration $100$ times 
and the final fraction of `$-$' agents is between $0.48-0.52$ in all the simulations. Panel 
(a) shows the initial population of the agents ($t=0$), black (resp. white) denotes `$-$' 
(resp. `$+$') agents. In panel (b), all agents reach the threshold by $t=2657346$. 

Since the final fractions of `$+$' and `$-$' agents are almost the same, the basic difference 
between panels (a) and (b) is the rearrangements of the agents into clusters. 
Panels (c) and (d) in Fig.\ref{fig1} are for a simulation when the initial population of `$-$' agents is $0.7$ of the 
total. Cluster formation is more pronounced in this simulation. The final fraction of `$-$' agents in panel (d) 
is $0.9$                                              
with a variation of $\pm 0.02$ for $100$ simulations. We note here that the behavior of our model has some similarity 
with the nonlinear voter model \cite{SB} in this aspect of cluster formation. 

Fig.\ref{figg1} shows the conversion of agents from `$-$' to `$+$' and vice versa. Panel (a) in Fig.\ref{figg1} 
shows this conversion for the simulations in Fig.\ref{fig1} on a $256\times 256$ lattice. 
We have plotted these graphs by counting agents that had an initial `$-$' opinion, but had $\theta^+>\theta^-$ at a particular simulation step, or vice versa.
The error bars are very small
compared to the values of the data points, hence they are not shown.
When the initial populations are balanced, the  conversions are almost in equal numbers and the conversions stop at around $t=1500000$,
thereafter the agents remain either `$+$ or `$-$' and gradually reach their thresholds.
This implies a
rearrangement of the agents in distinct clusters similar to the first order phase transition in a
2D random-field Ising model \cite{C}. 

When the simulation starts with an initial fraction $0.7$ of `$-$' agents, the final fraction of `$-$' agents is $0.9$
with a variation of $\pm 0.02$ for $100$ simulations. 
The conversion of `$+$' to `$-$' agents in this case is much higher and
the fraction of `$-$' agents increases from a fraction of $0.7$ to over $0.9$. However, still there are some
conversions from `$-$' to `$+$' agents. The final configuaration at $t=2500178$ shows
islands of `$+$' agents due to strong surface tension.

Panel (b) in Fig.\ref{figg1} shows similar conversions of agents for a simulation on a completely 
connected network (CCN) of size $65536$. These simulations are very sensisitive to the value of $\tau$. 
Both the simulations have an initial fraction of `$-$' agents as $0.5$, increasing the value of $\tau$ quickly pushes the 
final population to either all `$-$' or all `$+$' agents and this final population differs from simulation to simulation. 
The final population is an equal mix of `$-$' and `$+$' agents for $\tau=3$.  

The formation of clusters on the lattice is symptomatic in our model, as there 
is strong surface tension along the boundaries between regions of `$-$' and `$+$' opinions. 
We also exerimentally verified the surface tension in our model 
by seeding a lattice of size $256\times 256$ with a droplet of negative opinion and let the system evolve until 
all the agents reach their thresholds, as shown in Fig. \ref{fig2}. There was almost no change in the shape of the droplet
except for minute changes on the boundary. This is  different from the 
voter model, as the coarsening of a similar droplet under the voter model results 
due to lack of surface tension and the droplet disintegrates into a region with an 
irregular boundary \cite{DCCH}.    

\begin{figure}
\centering
\includegraphics[width=\columnwidth, height=4cm]{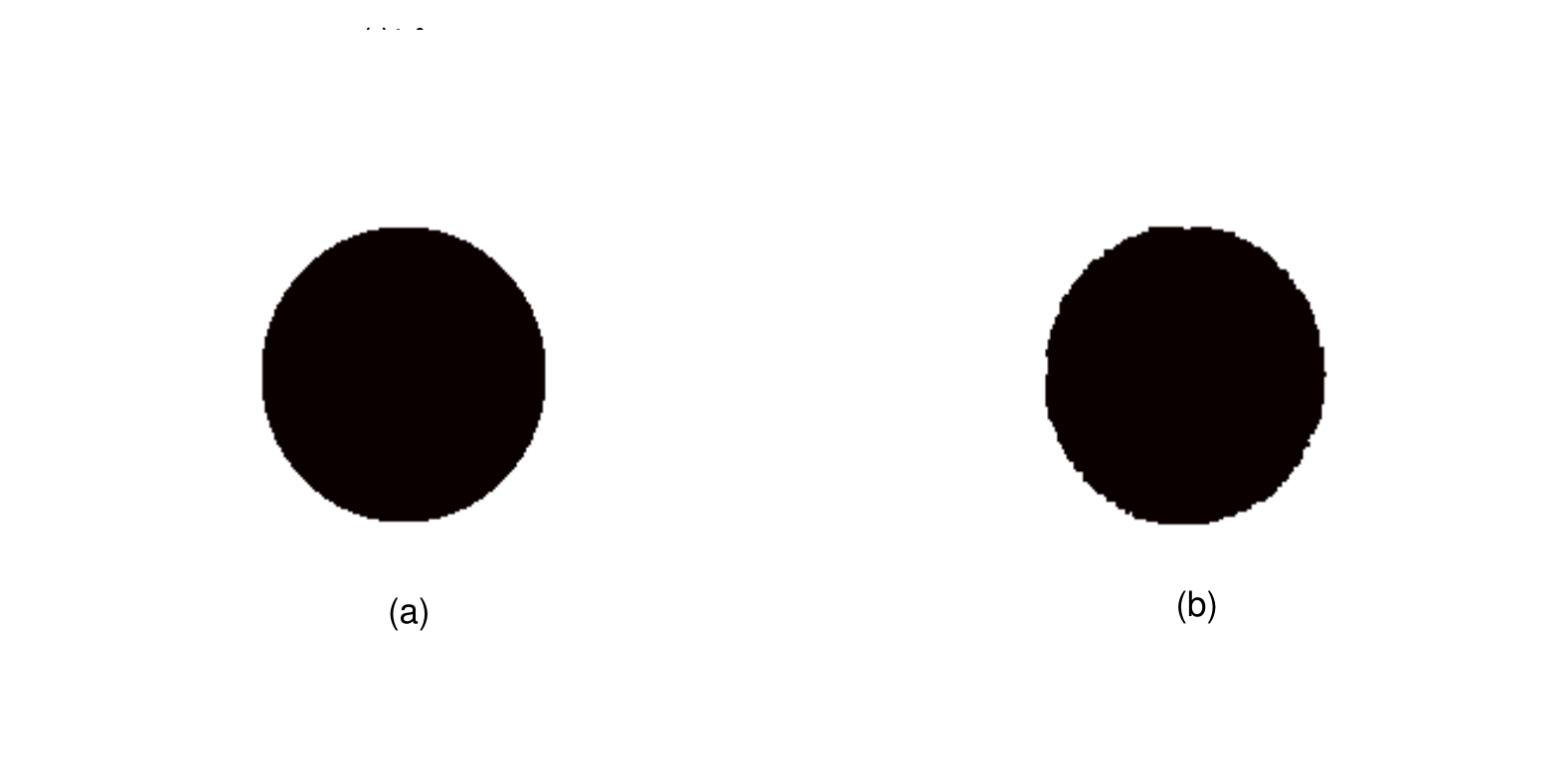}
\caption{Time evolution of a droplet of `$-$' agents in our model. {\bf (a)} t=0; {\bf (b)} t=1985384.}
\label{fig2}
\end{figure}

The effects of threshold in our model for a completely connected network and for a lattice 
are quite different. Increasing the threshold for the lattice has a much slower effect. 
This we can again attribute to the strong surface tension in our model. The formation of the 
clusters or islands of `$-$' agents is fairly rapid irrespective of the threhsold and the 
main effect of the threshold is the increase in convergence time when all the agents reach 
their thresholds within the clusters. On the other hand the model converges to an all 
`$-$' or all `$+$' population with higher threshold for the completely connected network. 

We studied the formation of clusters of `$+$' and `$-$' agents on a lattice.  
The clusters of `$+$' and `$-$' agents are of similar size when the  
starting population of `$+$' and `$-$' agents is $0.5$ of the total population each. The internal sites of 
clusters have all their neighbors as `$+$' or `$-$' sites, whereas the sites on the surface of 
clusters have a mixed number of `$+$' and `$-$' neighbors. In Fig.\ref{neighbors}(a) we study the change in 
the population of lattice points with different numbers of `$+$' and `$-$' neighbors 
on a lattice of size $1024\times 1024$. The 
trends for both `$+$' and `$-$' neighbors are similar. In Fig.\ref{neighbors}(a), the number of `$+$' sites with a single
`$-$' neighbor grows very fast and stabilizes at a high level as more and more lattice sites become
parts of larger clusters. These `$+$' sites with a single `$-$' neighbor are on the surface or the boundary of 
the clusters. On the other hand the number of `$+$' sites with two to four `$-$' neighbors decrese rapidly 
 and stabilize at lower levels, as these sites become parts of clusters. 

We study a ratio $\phi$ in Fig.\ref{neighbors}(b). This is the ratio of lattice sites on the cluster boundaries 
(sites that have neighbors of opposite opinion) and the total number of possible neighbors in the entire 
lattice. We have plotted three graphs with starting populations of `$-$' agents as $0.1$, $0.2$ and $0.4$, 
with varying target populations of `$-$' agents (a fraction of $0.1$ higher than the starting population, until a 
fraction of $0.9$). The graphs have a general trend that $\phi$ decreases with an increase in the target fraction, 
as the clusters of `$+$' agents decrease in size. However there is a slight increase in $\phi$ as the starting population 
of `$-$' agents is increased. This is due to formation of a higher number of `$-$' clusters as a higher initial 
population provides a larger number of seeds for these clusters.  

\begin{figure}
\centering
\includegraphics[width=\columnwidth,height=10cm]{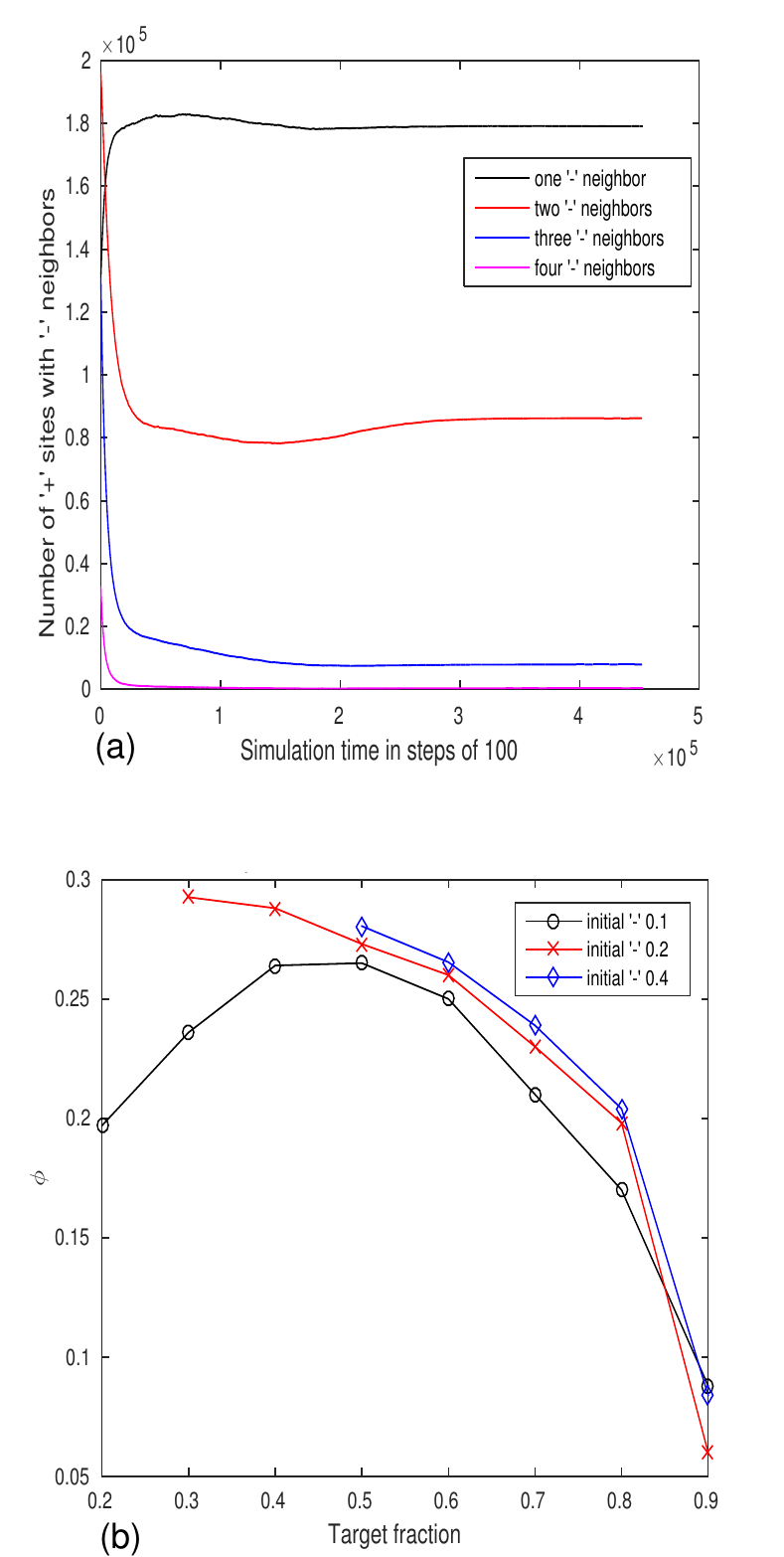}
\caption{{\bf (a)} A plot of change of numbers of `$+$' and `$-$' neighbors of lattice sites for a simulation on a 
lattice of size $1024\times 1024$. The simulation has been done only partially until the clusters stabilize. The top line shows a rapid increase of `$+$' sites that have one `$-$' neighbor. The number of `$+$' sites with two, three and four `$-$' neighbors (second from top to bottom) decrease. 
{\bf (b)} $\phi$ against target fraction for three different initial populations of `$-$' agents.}
\label{neighbors}
\end{figure}

As the `$+$' and `$-$' agents form clusters quite rapidly in simulations on a lattice, 
it is natural that most of the agents will be inside the clusters and a relatively small number of 
agents will be on the surface or the boundary of the clusters. This behavior of our model is quite similar to 
the random-field Ising model \cite{CN} in this respect. Moreover there is a  power-law relationship between the 
lattice points inside the clusters and on the surface of the clusters. If we denote the number of lattice sites 
inside the clusters (resp. on the surface) as $V$ (resp. $S$), this relationship can be 
expressed as 

\begin{equation}
\label{eqnpower}
S=cV^{\delta}
\end{equation} 

\begin{figure}
\centering
\includegraphics[width=\columnwidth,height=6cm]{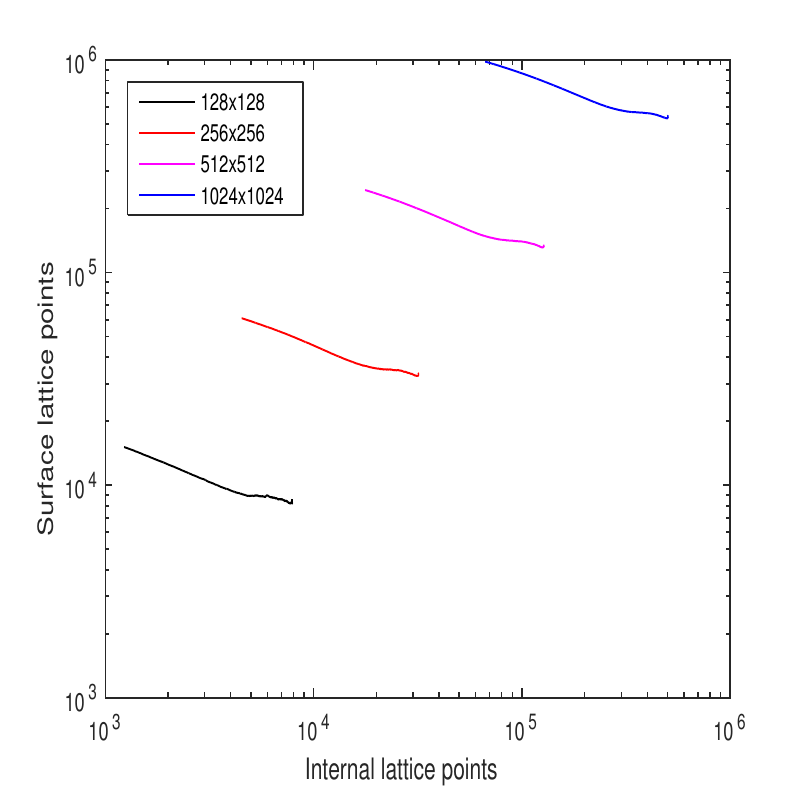}
\caption{A log-log plot of surface versus internal points in clusters for different sizes of lattice. The size of the lattice increases 
from $128\times 128$ (bottom) to $1024\times 1024$ (top). }
\label{power}
\end{figure}

We show the log-log plot of this equation in Fig. \ref{power} for different sizes of lattice.
A data collapse shows that the exponent 
$\delta=-0.29$ in case of the simulation with initial fraction of `$+$' and `$-$' agents $0.5$ each. 
The exponent $\delta$ depends on the initial population of `$+$' and `$-$' agents. For example, 
with a simulation starting with $0.3$ `$-$' and $0.7$ `$+$' agents, the final population 
of `$+$' agents is $0.8\pm 0.02$ and $\delta=-0.59$. This is due to the fact that the number 
of clusters of `$-$' agents as well as the sizes of the clusters are smaller in this case and 
hence the lattice sites on the cluster surfaces are also much smaller in number.

\section{IV. Dynamics with bias}

Our aim in this section is to use a power-law bias to increase the population of `$-$' 
agents when the starting population of `$-$' agents is $<0.5$ and the target population of 
`$-$' agents is higher than the starting population. 
As we have noted in the previous section the conversion of agents from `$+$' to `$-$' and 
vice versa is rapid in the early stages of the simulation both for the completely connected network 
and the lattice. The purpose of introducing the power-law bias in eqn.(\ref{eqn1}) is to 
influence this conversion so that a larger proportion of `$+$' agents convert to `$-$' opinion. 
In eqn.(\ref{eqn1}), $0\leq p_3\leq 1$ and $\beta\geq 0$, and we have discussed the case $\beta=0$ 
in the previous section. Hence for a fixed $p_3$, $p_3^\beta$ is a monotonically decreasing function 
of $\beta$. The condition  $\theta^-_i> p_3^{\beta}\theta^+_i$ in eqn.(\ref{eqn1}) ensures that 
this condition will be satisfied for lower values of $\theta^-_i$ compared to $\theta^+_i$, as the 
factor $p_3^\beta<1$ and  reduces the magnitude of the right hand side in eqn.(\ref{eqn1}).  

We consider the $\tau$ discrete integer values of $\theta^-_i$, i.e., $\theta^{-}_{i,1}, \theta^{-}_{i,2}, 
\dots, \theta^{-}_{i,\tau}$ for a fixed threshold $\tau$. Similarly we consider the $\tau$ discrete 
integer values of $\theta^+_i$, namely, $\theta^{+}_{i,1}, \theta^{+}_{i,2}, 
\dots, \theta^{+}_{i,\tau}$. The effect of the factor $p_3^\beta$ on $\theta^+_i$ is a mapping 
$\theta^+_i\rightarrow \theta^-_i$ to partition the 
values $\theta^{+}_{i,1}, \theta^{+}_{i,2},  \dots, \theta^{+}_{i,\tau}$ into $k$ partitions $P_k, 1\leq k\leq \tau$.
The members of partition $P_m$ are mapped within 
two consecutive integer values in $\theta^-_i$. 
For example, if we assume $\tau=10, p_3=0.9$, and $\beta=2.6$, $p_3^\beta=0.76$. There are $10$ values each for 
$\theta^-_i$ and $\theta^+_i$, the integers $1,2,\dots, 10$. Hence $p_3^\beta\theta^+_i$ can be partitioned into 
eight partitions that are within consecutive integer values of $\theta^-_i$, $[0,1]_{[1]}$, $[1,2]_{[2]}$, $[2,3]_{[3]}$, $[3,4]_{[4,5]}$, $[4,5]_{[6]}$, 
$[5,6]_{[7]}$, $[6,7]_{[8,9]}$, $[7,8]_{[10]}$. For example, $[6,7]_{[8,9]}$ indicates that $p_3^\beta\theta^+_i$ is between 
$6$ and $7$ for $\theta^+_i=8,9$ ($0.76\times8=6.08$ and $0.76\times9=6.84$).
The dynamics of our model remains unaffected for different values of $\beta$ that result in the same partitions, as the 
condition in eqn.(\ref{eqn1}) will evaluate identically for the same partition. 
In this example  $\beta=2.7$ also gives the same partitions  for $p_3^\beta\theta^+_i$ as $\beta=2.6$, and hence the 
behavior of our model will remain the same for either of these two choices for $\beta$. Consequently the 
exponent $\beta$ has a range instead of a unique value for achieving a final population of `$-$' agents and
there is no change in the final population when the $\beta$ value remains within this range. 
However the changes in the final population are sharp whenever the $\beta$ parameter causes a transition from 
one partition to another.

When the groups of $p_3^\beta$ are compressed within lower values of $\theta^-_i$, the result is an 
increase of the $\theta^-_j$ counter as the condition in eqn.(\ref{eqn1}) is satisfied even for 
lower values of $\theta^-_i$. As a result this favors the `$-$' agents to dominate the dynamics as 
more and more agents (both with initial `$+$' or `$-$' opinions) reach their thresholds for the 
$\theta^-$ counters. Hence a high enough threshold makes  the system to converge in  an all `$-$' opinion scenario. 
This convergence is faster for the completely connected network compared to the lattice, as the erosion 
of the surface of the clusters of `$+$' nodes is a much slower process for the lattice.

\subsection{A. Dynamics on a completely connected network}

We first study the dynamics of the system through an example for the completely connected network 
when the initial fraction of `$-$' (resp. `$+$') agents is 
$0.3$ (resp. $0.7$) of the total population and the required final fraction of `$-$' agents is $0.9$ of the 
total population. 
As we have noted earlier, if we simulate this scenario without a bias, i.e., when $\beta=0$, the final fraction 
of agents with `$-$' opinion reduces further. Also, the final fraction depends on the threshold, 
it approaches $0$ very fast as the threshold is increased for the completely connected network. 
On the other hand a low threshold does not allow enough scope for the conversion of a 
large number of `$+$' agents to `$-$' agents that is required for achieving a high fraction of `$-$' agents starting 
from a low fraction. This trade-off for the threshold exists for the model with a bias as well. In this case the 
bias gives an impetus for reversing `$+$' agents to `$-$' agents. If the threshold is high, eventually all `$+$' 
agents will be converted and the final population will consist of all `$-$' agents. 
 
\begin{figure}
\begin{center}
\includegraphics[width=\columnwidth, height=9cm]{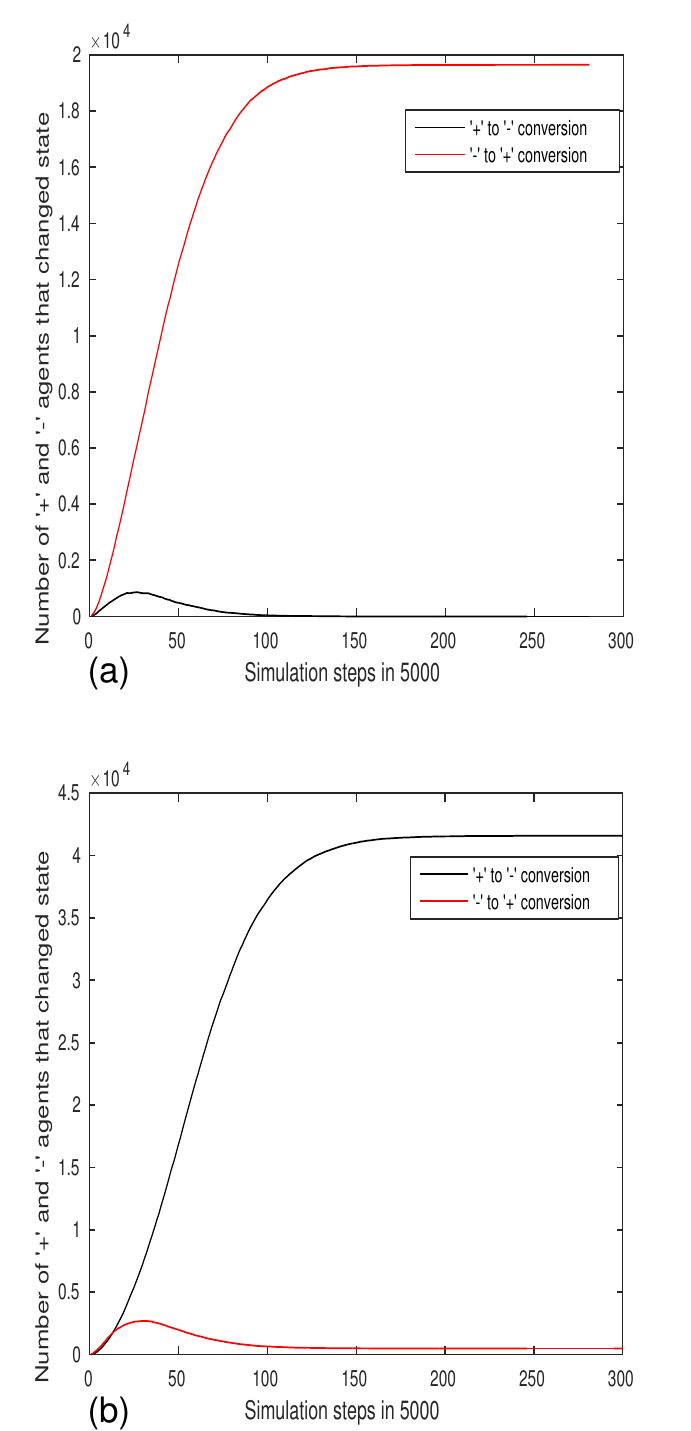}
\end{center}
\caption{The conversion of agents from `$+$' to `$-$' and `$-$' to `$+$' for the biased and unbiased models for 
a simulation of $256\times 256=65536$ agents on a completely connected network. $\tau=5$ and 
$p_3=0.9$. The starting population of `$-$' agents is $0.3$ of the total in both cases. 
If $\theta^-_i>\theta^+_i$, agent
$i$ is identified as a `$-$' agent, and vice versa.
{\bf (a)} $\beta=0$ and almost all `$-$' agents are converted to `$+$' agents (the upper curve), conversion of `$+$' to `$-$' is very low (lower curve). 
{\bf (b)} $\beta=6.7$ and the bias forces conversion of a large number of `$+$' agents to `$-$' (upper curve), compared to `$-$' to `$+$' conversion
(lower curve).} 
\label{bias-complete}
\end{figure} 
  
Fig. \ref{bias-complete} shows the comparison between the simulations of our model with and without bias. 
This simulation has been done on a completely connected network of $65536$ agents. The starting population 
of `$-$' agents is a fraction $0.3$ of the total, and the final target population of `$-$' agents is a fraction of $0.9$ of 
the total population. The threshold $\tau=5$ in this case. The simulation has been done without bias in {\bf (a)}, and 
a lower starting population of `$-$' agents drives the system to a final population of all `$+$' agents. 
We show in Fig.\ref{bias-complete}(a) the conversion of agents from `$+$' to `$-$' and vice versa. 
There is a small initial conversion of `$+$' agents to `$-$', however, soon the larger population of `$+$' agents dominate 
and all of the initial fraction of $0.3$ of `$-$' agents convert to `$+$'. The graph shows a cumulative number of the agents 
that have higher value in the $\theta^+$ or $\theta^-$ counter  until a particular step. Almost all the conversions occur in the initial stages of the simulation and 
the agents reach their thresholds afterwards over many simulation steps (as indicated by the red curve in (a)). 
The biased simulation is shown in (b). The bias in this case drives a conversion of `$+$' agents to `$-$', 
and as a result the final population of `$-$' agents rise to a fraction of $0.9\pm 0.03$ for $100$ simulations. 

The threshold for both the simulations is $\tau=5$. Increasing this threshold for the simulation with bias 
quickly pushes the final population to consist only of `$-$' agents as a higher threshold gives more scope for 
`$-$' agents to convert `$+$' agents. For example a simulation with $\tau=10$ has a final population of all `$-$' agents. 
As $\beta=6.7$ in the simulations in (b), and the final target fraction of `$-$' agents is $p_3=0.9$, the bias factor 
$p_3^\beta=0.9^{6.7}=0.493$. Since $\tau=5$, there are three partitions of $p_3^\beta\theta^+_i$ for agent $i$. These are 
$[0,1]_{[1,2]}$, $[1,2]_{[3,4]}$ and $[2,3]_{[5]}$. Most of the transitions from `$+$' to `$-$' occur in the initial 
steps of the simulation as shown in Fig. \ref{bias-complete}. There is a small number of conversions from `$-$' to `$+$' as
well, however, both of these conversions plateau realtively early in the simulation. Another interesting aspect of the 
dynamics is that a complete  coversion of `$+$' to `$-$' agents is dependent on $\tau$, rather than $\beta$, as mentioned 
earlier. For example, 
$\beta=18$ gives only one partition of $p_3^\beta\theta^+_i$, $[0-1]_{[1-5]}$. However, simulations in this case show 
a final population of `$-$' agents as a fraction $0.95\pm0.02$ of the total population. This is only possible if some of the `$+$' 
agents meet only `$+$' agents during the course of the simulation, as a `$+$' agent will be converted to a `$-$' agent 
whenever a `$+$' agents meets a `$-$' agent in this case. Table \ref{table1} shows some more results from our simulations. 

\begin{table}
\centering
\begin{tabular}{ |c|c|c|c|}
\hline
initial population & $\beta$ & target population & population achieved \\
\hline
0.1 & 3.1 & 0.6 & 0.59-0.63 \\ \hline
0.1 & 3.8 & 0.7 & 0.68-0.73 \\ \hline
0.2 & 1.4 & 0.6 & 0.59-0.65 \\ \hline
0.2 & 3.0 & 0.7 & 0.69-0.73 \\ \hline
0.4 & 0.6 & 0.7 & 0.67-0.73 \\ \hline 
0.4 & 1.8 & 0.8 & 0.79-0.83 \\ \hline
\end{tabular}
\caption{Simulation results on a completely connected network of $256\times 256=65536$ agents. The result  
in each row is collected from $100$ simulations. The threshold is $\tau=5$ in all cases. The results were similar  
when simulations were run $10$ times each on completely connected networks  of size $512\times 512=262144$ and 
$1024\times 1024=1048576$ agents. }
\label{table1}
\end{table}

\begin{figure*}[t]
\centering
\includegraphics[width=\textwidth, height=6cm]{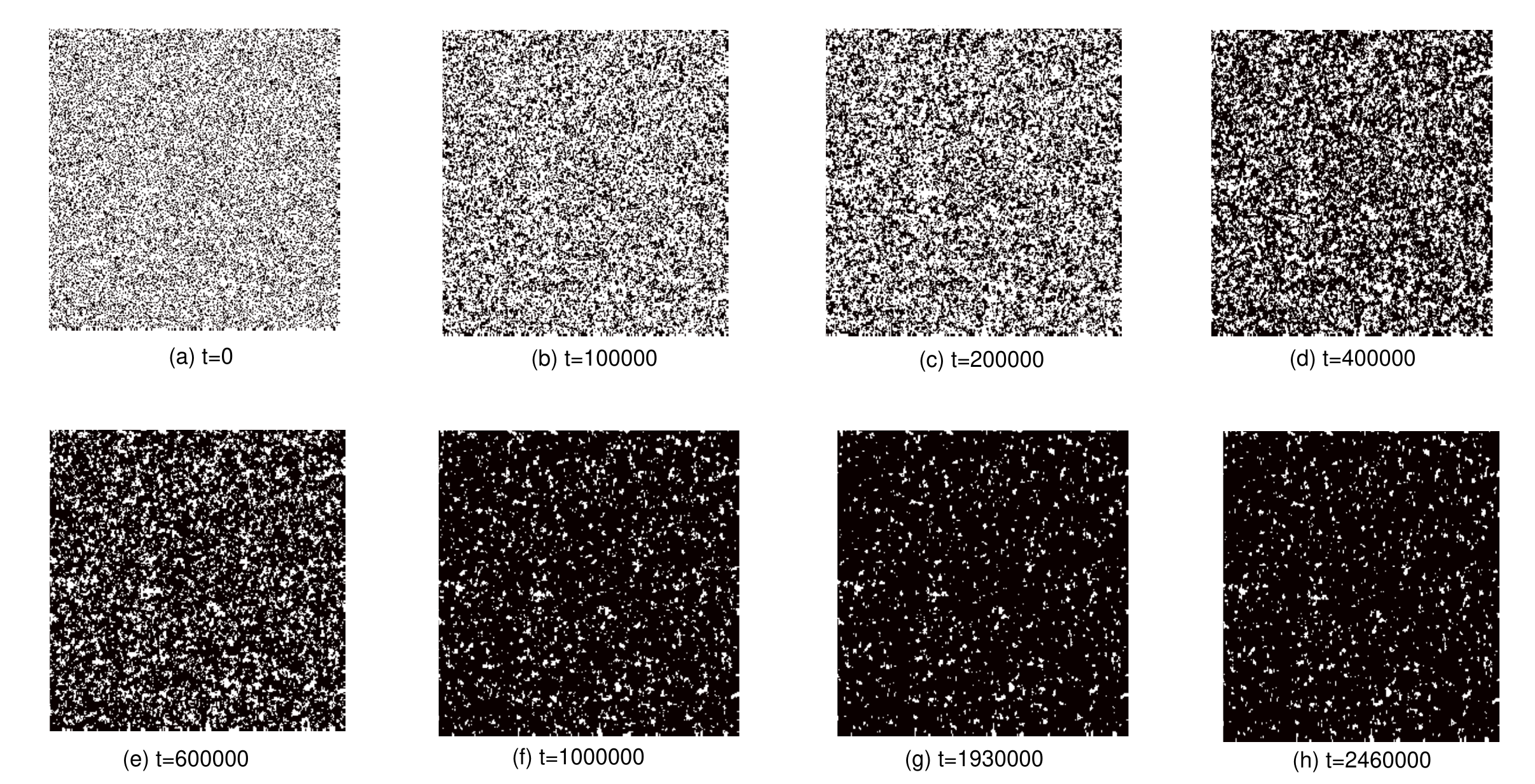}
\caption{Evolution of agents on a $256\times 256$ lattice, with initial
population of negative agents is a fraction $0.3$ of the total. `$-$' agents are shown in black and `$+$' 
agents in white. 
The data was collected over $50$ simulations and the final population of negative agents
was $0.89-0.93$ of the total. This evolution is shown at different time steps for a representative
simulation.}
\label{lattice-bias}
\end{figure*}

\subsection{B. Dynamics on a lattice}

We now discuss the dynamics of the system on a lattice when $\beta$ is non-zero in eqn.(\ref{eqn1}). 
We take the same representative case when the initial population of `$-$' agents is a fraction  $0.3$ of the total and
the final desired population of `$-$' agents is a fraction $0.9$ of the total. For the simulation on 
a $256\times 256$ lattice, we have used $\beta=10.6$ and obtained a final population of `$-$' agents between a fraction 
$0.89-0.93$ for $50$ simulations. We have chosen $\tau=10$, as a higher threshold has a slower effect in driving the system to an 
all `$-$' population, compared to the simulations on completely connected networks. 
A representative simulation is shown in Fig. \ref{lattice-bias}. Panel (a) shows the initial configuration 
with `$-$' and `$+$' agents $0.3$ and $0.7$ of the total population. Panel (b) shows the initial growth 
of the number of `$-$' agents with the initial `$-$' agents as seeds. The number of `$-$' agents has grown 
significanltly even after $t=100000$ Monte Carlo steps. Panels (c) and (d) show the simulation at time steps 
$t=200000$ and $t=400000$ respectively and  the growth of the clusters of `$-$' agents is clearly visible. Panel (e) shows 
the simulation at $t=600000$ and the large clusters of `$-$' agents have already emerged. These clusters further 
consolidate in panel (f) at $t=1000000$ and remain almost unchanged until the end of the simulation 
at $t=2469567$. This is easy to see from panels (f),(g) and (h). 

\begin{figure}
\centering
\includegraphics[width=\columnwidth, height=9cm]{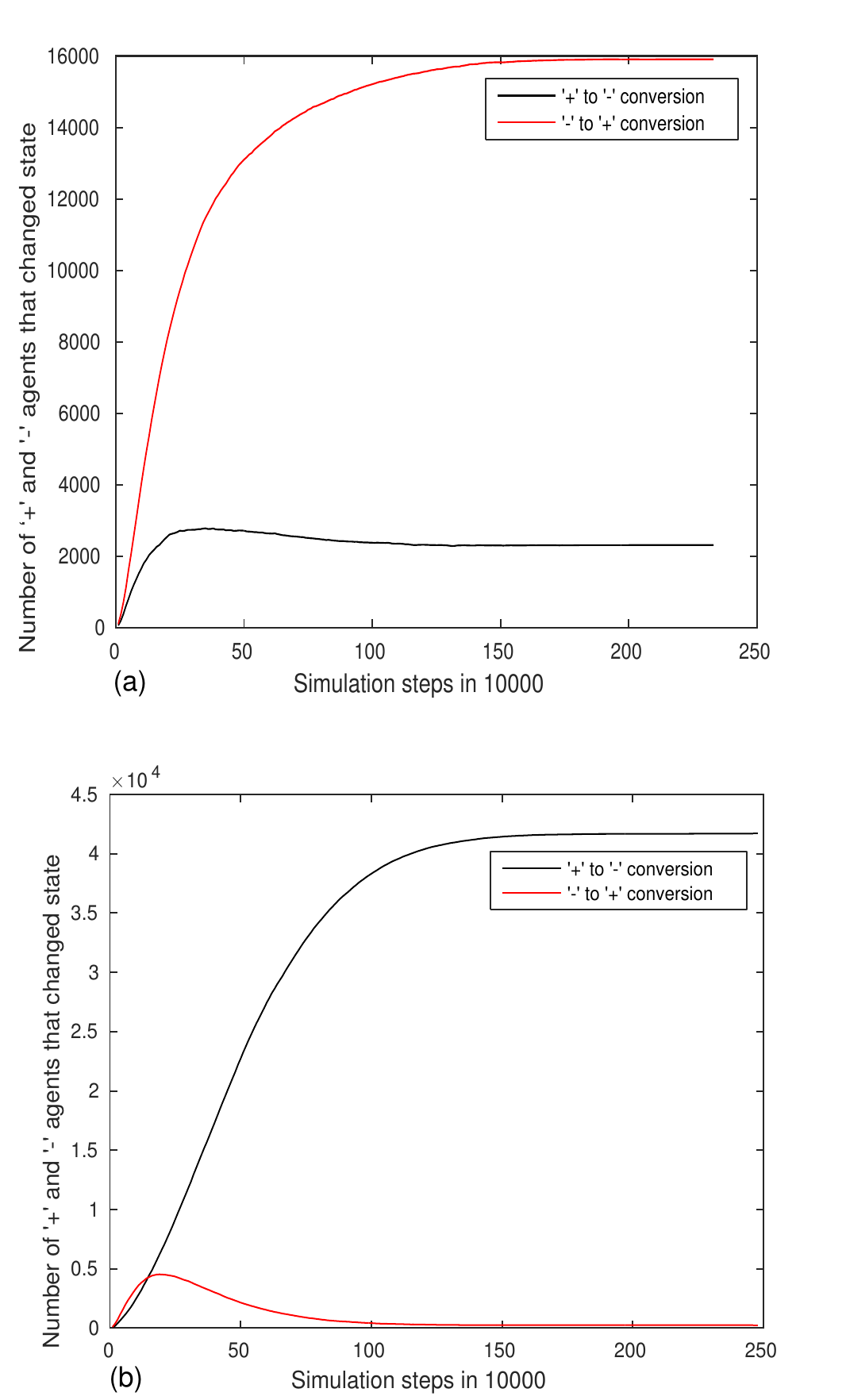}
\caption{The conversion of agents from `$+$' to `$-$' and vice versa for a simulation on a $256\times 256$ 
lattice, with initial fraction of `$-$' agents $0.3$, $p_3=0.9$ and $\tau=10$. {\bf (a)} $\beta=0$, almost all `$-$' agents are converted to `$+$' (upper curve), compared to `$+$' to `$-$' conversions (lower curve); {\bf (b)} $\beta=10.6$, the bias forces the conversion of a large number of `$+$' agents to `$-$' (upper curve) compared to `$-$' to `$+$' conversion (lower curve).}
\label{lattice-conversion}
\end{figure}
We show in Fig. \ref{lattice-conversion} the conversion of agents from  `$-$' to `$+$' and vice versa. 
The behavior is similar to the simulation on the lattice 
as the large-scale conversion of `$+$' to `$-$' agents occur quite early in the simulation. However, 
there is more conversion of `$-$' agents to `$+$' agents initially on the lattice compared to the 
completely connected network. 
The simulation takes a 
long time to complete since the agents reach their thresholds $\tau=10$ at the later stages of the simulation. 
The fractions of final population of `$-$' agents for lattices of size $512\times 512$ and $1024\times 1024$ are 
within these bounds when $\beta=10.6$ and $\tau=10$ are used for the simulations.  
The partitions of $\theta^+$ values with $\tau=10$ and $\beta=10.6$ are $[0,1]_{[1-3]}$, $[1-2]_{[4-6]}$,$[2-3]_{[7-9]}$, 
$[3-4]_{[10]}$. It is evident that the dynamics of the system is dominated by the partitions $[0,1]_{[1-3]}$, $[1-2]_{[4-6]}$ as
the conversions of `$+$' to `$-$' agents are rapid in the early stages of the simulation when the $\theta^+$ and 
$\theta^-$ counters of all agents have relatively lower values. This is similar to the simulations on the completely connected network.   

\begin{table}
\centering
\begin{tabular}{ |c|c|c|c|}
\hline
initial population & $\beta$ & target population & population achieved \\
\hline
0.1 & 3.0 & 0.6 & 0.59-0.64 \\ \hline
0.1 & 5.2 & 0.7 & 0.68-0.73 \\ \hline
0.2 & 1.8 & 0.6 & 0.59-0.64 \\ \hline
0.2 & 3.0 & 0.7 & 0.65-0.77 \\ \hline
0.4 & 1.2 & 0.7 & 0.68-0.73 \\ \hline
0.4 & 2.8 & 0.8 & 0.76-0.83 \\ \hline
\end{tabular}
\caption{Simulation results on a lattice  of $256\times 256$ agents. The result
in each row is collected from $100$ simulations. The threshold is $\tau=10$ in all cases. The results were similar
when simulations were run $10$ times each on  lattices of size $512\times 512$ and
$1024\times 1024$ agents. }
\label{table2}
\end{table}

Table \ref{table2} shows some more results from our simulations. 
We should note that it is possible to use  higher values of $\tau$ for achieving 
sharper and more stable population fractions closer to the target population. We illustrate this in Fig. \ref{figx} with $0.3$ as the starting fraction of 
`$-$' agents and $0.8$ as the target fraction. We vary $\tau$ for two fixed values of $\beta$. For $\beta=4.6$, the target fraction is reached 
at a lower value of $\tau=9$, however average target fraction was $0.82$ for $20$ simulations on a $256\times 256$ lattice. On the 
other hand a lower value of $\beta=2.3$ results in a slow convergence to the target fraction of $0.8$ at $\tau=25$, and the 
target fraction was $0.8$ every time for $20$ simulations. This behavior is common in our simulations, and its is possible to choose $\beta$ 
and $\tau$ pairs that allow us to achieve the target fraction accurately. 

Fig. \ref{figx} has some similarities with rate-distortion curves studied in information theory \cite{CT}. The aim 
of rate-distortion theory is to establish a connection between the channel capacity (rate)  and output performance 
(distortion) of a communication 
channel, through minimizing channel distortion captured through a cost function. A rate-distortion curve separates 
the plane into two regions, allowable and non-allowable. The points in the allowable region indicate the minimum 
required rate to achieve a particular distortion in the output signal. Points in the non-allowable region 
indicate distortions that are unachievable using the corresponding rates. Two extreme points on a rate-distortion curve 
are the minimum rate required for zero distortion and the maximum distortion when the rate is zero. This also 
indicates a trade-off between the channel capacity and distortion, as distortion reduces by increasing 
channel capacity and increases by reducing channel capacity.

We can draw a parallel of Fig. \ref{figx} with a rate-distortion curve if we consider the interactions of agents 
in eqns. (\ref{eqn1}) and (\ref{eqn2}) as the channel capacity or rate, and the fraction $\phi$ as the output of the channel. 
Increasing $\tau$ increases the number of interactions between agents and can be seen as an increase in channel 
capacity. The distortion is the difference between the fraction $\phi$ achieved with a specific value of $\tau$ and 
the target fraction of `$-$' agents. 
We can study a trade-off between $\tau$ and $\phi$ for a fixed $\beta$. For example, for $\beta=2.3$ (the red line marked with `x'
in Fig. \ref{figx}),  if we fix a value of $\tau$ and draw a vertical line, all the fractions $\phi$ of final 
population of `$-$' agents below the red line are achievable with 
$\beta=2.3$. And no fraction $\phi$ above the red line is achievable with $\beta=2.3$. In other words, if we fix $\beta$, the 
red line divides the plane into allowable (below) and non-allowable (above) regions. A trade-off between $\tau$ and 
$\phi$ is also noticeable, as the non-allowable region is larger with smaller values of $\tau$ and vice versa. Similar 
trade-off due to rate-distortion curves has been observed in diverse domains like human 
perception \cite{S}, capital asset pricing model for stocks \cite{O} and balance between growth and entropy in 
bacterial cultures \cite{MCM}.

\begin{figure}
\centering
\includegraphics[width=\columnwidth,height=5cm]{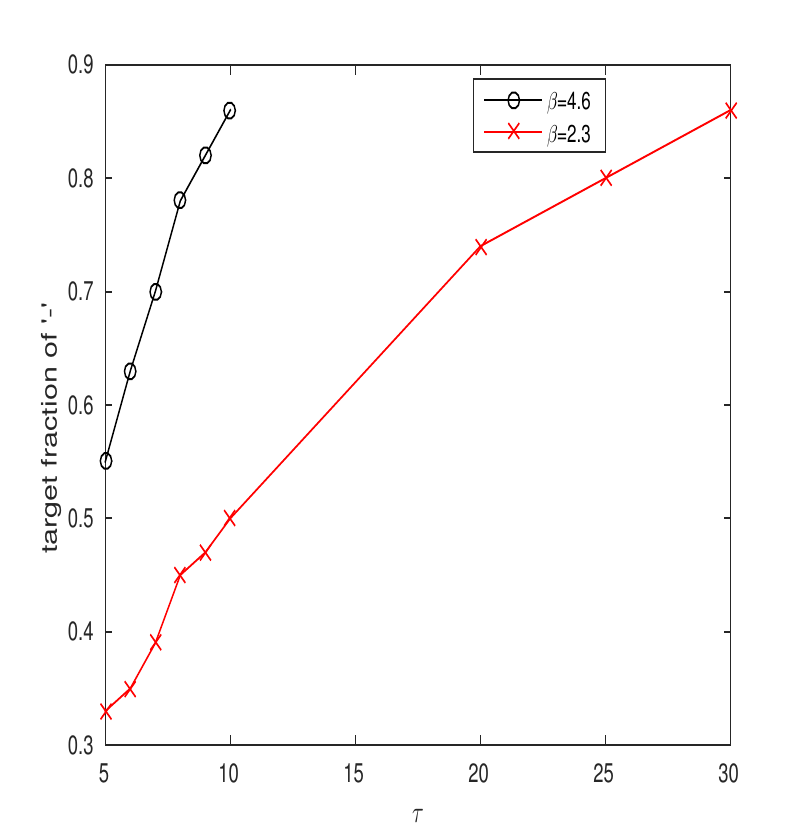}
\caption{\small Slow and accurate convergence to the target fraction $0.8$ with 
different values of $\beta$ and $\tau$. }

\label{figx}
\end{figure}

\begin{table}
\centering
\begin{tabular}{ |c|c|c|c|}
\hline
initial population & $\beta$ & target population & achieved \\
\hline
0.1 & $3.5\pm 2.0$ & 0.6 & 0.59-0.63 \\ \hline
0.1 & $5.5\pm 2.0$ & 0.7 & 0.69-0.73 \\ \hline
0.2 & $1.8\pm 1.1$ & 0.6 & 0.59-0.63 \\ \hline
0.2 & $3.0\pm 2.0$ & 0.7 & 0.69-0.73 \\ \hline
\end{tabular}
\caption{Simulation results on a completely connected network   of $256\times 256$ agents. The result
in each row is collected from $100$ simulations. The threshold is $\tau=5$ in all cases. The results were similar
when simulations were run $10$ times each on completely connected networks of $512\times 512$ and
$1024\times 1024$ agents. }
\label{table3}
\end{table}

\subsection{C. Dynamics with a faulty $\beta$}

We have also experimented with the dynamics of the system when the bias $\beta$ is not constant
for all the agents, rather $\beta$ varies within a range of values that we
denote by $\beta\pm R$. We choose a value for $\beta$ from the range
$\beta-R$ to $\beta +R$ uniformly at random at each Monte Carlo step for each agent. In other words, 
each agent uses a different $\beta$ within this range at each Monte Carlo step. Though
the values of $\beta$ are different for achieving the desired fractions of
final population of `$-$' agents, the system is stable for a range of $\beta$ that is $\pm 2.0$ around a central value of $\beta$ for the cases when the
central value of $\beta>2.0$. The deterioration in achieving the final desired
fraction of `$-$' agents starts beyond the $\pm 2.0$ range. Some results are shown in Table \ref{table3} for a
completely connected network, and in Table \ref{table4} for lattices. The dynamics is quite similar in both cases.
We have also verified that these results scale for larger networks.

\begin{table}
\centering
\begin{tabular}{ |c|c|c|c|}
\hline
initial population & $\beta$ & target population & achieved \\
\hline
0.1 & $3.2\pm 2.0$ & 0.6 & 0.59-0.63 \\ \hline
0.1 & $5.2\pm 2.0$ & 0.7 & 0.69-0.72 \\ \hline
0.2 & $3.0\pm 2.0$ & 0.7 & 0.69-0.73 \\ \hline
0.4 & $2.4\pm 2.0$ & 0.8 & 0.79-0.83 \\ \hline
\end{tabular}
\caption{Simulation results on a lattice  of $256\times 256$ agents. The result
in each row is collected from $100$ simulations. The threshold is $\tau=10$ in all cases. The results were similar
when simulations were run $10$ times each on lattices of size $512\times 512$ and
$1024\times 1024$ agents. }
\label{table4}
\end{table}

\section{V. Discussion}

We have presented a model of opinion dynamics based on the negativity bias extensively studied in the 
psychology literature. 
Our main aim was to investigate the effect of negativity bias in binary opinion formation.  
One of the interesting aspects of our model is the formation of stable 
target population of `$-$' agents. 
Our model is close to real-world exchange of opinions based on negativity bias. 
People with different opinions usually discuss pros and cons of both alternatives and gives more importance to 
negative opinions. We have abstracted this 
real-world situation in terms of the two counters for individual agents. 
We have shown that application of a 
power-law bias during opinion exchange results in consistent target populations and the bias factors are scale-invariant. 
Moreover, we have also shown that this consistency is maintained with bias factors that can vary randomly and 
uniformly within a range. Another interesting aspect of our model is its rapid convergence, the composition of the final
population is reached quite early in the simulation, when each agent has interacted with only a few other agents. This
is again close to the real-world situation in the sense that usually people even within a large population 
interact with a few other people while making decisions. 

There are some similarities between the dynamics of our 
model and the dynamics of the 
random-field Ising model. For example, the conversions of `$+$' to `$-$' opinion 
and vice versa are similar to the first order phase transition in the 
random-field Ising model. Similarly, formation of clusters of `$+$' and `$-$' 
opinions and strong surface tension on cluster boundaries are very similar to 
the domains of similar spin in the random-field Ising model.  
We will explore these similarities further in future work.   

\section{acknowledgments}

The author thankfully acknowledges very important suggestions from three referees 
who helped him to precisely define the scope of this work and improve its technical presentation.

\end {document}